\documentclass[pra,aps,showkeys,groupedaddress,superscriptaddress,twocolumn,toc=flat]{revtex4-1}
\usepackage[colorlinks=true,linkcolor=blue,urlcolor=blue,citecolor=blue]{hyperref}

\usepackage[utf8x]{inputenc}
\usepackage{color}
\usepackage{bbm} 

\usepackage{amsfonts,amsmath,amssymb,stmaryrd}

\usepackage{graphicx}
\usepackage{subfigure}  
\usepackage{bbm} 
\usepackage{hyperref}
\usepackage{epsfig}
\usepackage{mathrsfs}
\usepackage{verbatim}
\usepackage{centernot}
\usepackage{ulem}
\usepackage{xspace}
\usepackage[english, status=draft]{fixme}       
\fxusetheme{color}                              %
\usepackage{nicefrac}

\renewcommand{\l}{\left(}
\renewcommand{\r}{\right)}

\newcommand{\bra}[1]{\langle#1|}
\newcommand{\ket}[1]{|#1\rangle}

\renewcommand{\ij}{{\langle \vec{i}, \vec{j} \rangle}}
\renewcommand{\H}{\hat{\mathcal{H}}}

\renewcommand{\c}{\hat{c}}
\newcommand{\f}{\hat{f}}
\newcommand{\fd}{\hat{f}^\dagger}

\newcommand{\cd}{\hat{c}^\dagger}

\newcommand{\hd}{\hat{h}^\dagger}
\newcommand{\h}{\hat{h}}

\newcommand{\hc}{\text{h.c.}}

\usepackage{array}

\usepackage{cancel,ifthen}
\newcommand{\cmnt}[2][NoInPuT]{\ifthenelse{\equal{#1}{NoInPuT}}{}{{\color{red}\sout{#1}}} {\color{blue} #2}}

\usepackage{bm}	
\renewcommand{\vec}[1]{\bm{#1}}

\begin{document}
\normalem	

\title{Microscopic spinon-chargon theory of magnetic polarons in the $t-J$ model}

\author{Fabian Grusdt}
\email[Corresponding author email: ]{fabian.grusdt@tum.de}
\affiliation{Department of Physics, Harvard University, Cambridge, Massachusetts 02138, USA}
\affiliation{Department of Physics and Institute for Advanced Study, Technical University of Munich, 85748 Garching, Germany}

\author{Annabelle Bohrdt}
\affiliation{Department of Physics and Institute for Advanced Study, Technical University of Munich, 85748 Garching, Germany}
\affiliation{Department of Physics, Harvard University, Cambridge, Massachusetts 02138, USA}

\author{Eugene Demler}
\affiliation{Department of Physics, Harvard University, Cambridge, Massachusetts 02138, USA}

\date{\today}

\begin{abstract}
The interplay of spin and charge degrees of freedom, introduced by doping mobile holes into a Mott insulator with strong anti-ferromagnetic (AFM) correlations, is at the heart of strongly correlated matter such as high-$T_c$ cuprate superconductors. Here we capture this interplay in the strong coupling regime and propose a trial wavefunction of mobile holes in an AFM. Our method provides a microscopic justification for a class of theories which describe doped holes moving in an AFM environment as meson-like bound states of spinons and chargons. We discuss a model of such bound states from the perspective of geometric strings, which describe a fluctuating lattice geometry introduced by the fast motion of the chargon. This is demonstrated to give rise to short-range hidden string order, signatures of which have recently been revealed by ultracold atom experiments. We present evidence for the existence of such short-range hidden string correlations also at zero temperature by performing numerical DMRG simulations. To test our microscopic approach, we calculate the ground state energy and dispersion relation of a hole in an AFM, as well as the magnetic polaron radius, and obtain good quantitative agreement with advanced numerical simulations at strong couplings. We discuss extensions of our analysis to systems without long range AFM order to systems with short-range magnetic correlations.
\end{abstract}


\maketitle

\section{Introduction and overview}
\label{secIntro}
Despite many years of intense research, key aspects of the phase diagram of the Fermi-Hubbard model remain poorly understood \cite{Lee2006,Keimer2015}. Simplifying variational theories are lacking in important parameter regimes, which makes the search for a unified field theory even more challenging. In other strongly correlated systems, microscopic approaches have proven indispensable in the development of field-theoretic descriptions. A prominent example is constituted by the composite-fermion theory of the fractional quantum Hall effect, which provides a non-perturbative but conceptually elegant explanation of a class of topologically ordered ground states \cite{JAIN1989,Read1989,Jain2007}. 

In the case of the 2D Fermi-Hubbard model at strong couplings, already the description of a single hole doped into a surrounding AFM represents a considerable challenge. This problem is at the heart of high-temperature superconductivity and strongly correlated quantum matter, where pronounced AFM correlations remain present at short distances even for relatively large doping. While many properties of a single hole moving in an AFM spin background are known, their derivation, in particular at strong couplings, requires sophisticated numerical simulations \cite{SchmittRink1988,Kane1989,Sachdev1989,Elser1990,Dagotto1990,Martinez1991,Liu1992,Boninsegni1992a,Boninsegni1992,Leung1995,Brunner2000,Mishchenko2001,White2001}. This includes such basic characteristics as the shape of the single-hole dispersion relation. 

One of the central obstacles in the search for a generally accepted theory of strongly correlated materials, and the rich phase diagram of high-temperature superconductors in particular, is the lack of a known unifying physical principle. Arguably one of the most influential approaches is the resonating-valence bond (RVB) paradigm suggested by Anderson \cite{Anderson1987}. While it yields satisfactory results at intermediate and large doping levels \cite{Lee2008a}, it is not powerful enough to accurately describe the low-doping regime, starting on the single-hole level, or capture the disappearance of antiferromagnetism observed upon doping. 

Recent experiments with ultracold fermions in optical lattices \cite{Boll2016,Cheuk2016,Parsons2016,Brown2017} suggest a new paradigm: By introducing short-ranged hidden string order, a connection has been demonstrated between the Fermi-Hubbard model at finite doping and an AFM parent state at half filling, both in 1D \cite{Hilker2017,Salomon2018} and recently also in 2D \cite{Mazurenko2017,Chiu2018}. While the 1D case can be rigorously proven \cite{Ogata1990,Zaanen2001,Kruis2004a}, it was argued that this hidden string order also emerges in 2D as an immediate consequence of the hole motion \cite{Bulaevskii1968,Brinkman1970,Trugman1988,Shraiman1988a,Manousakis2007,Golez2014,Grusdt2018PRX}. 

\begin{figure}[t!]
\centering
\epsfig{file=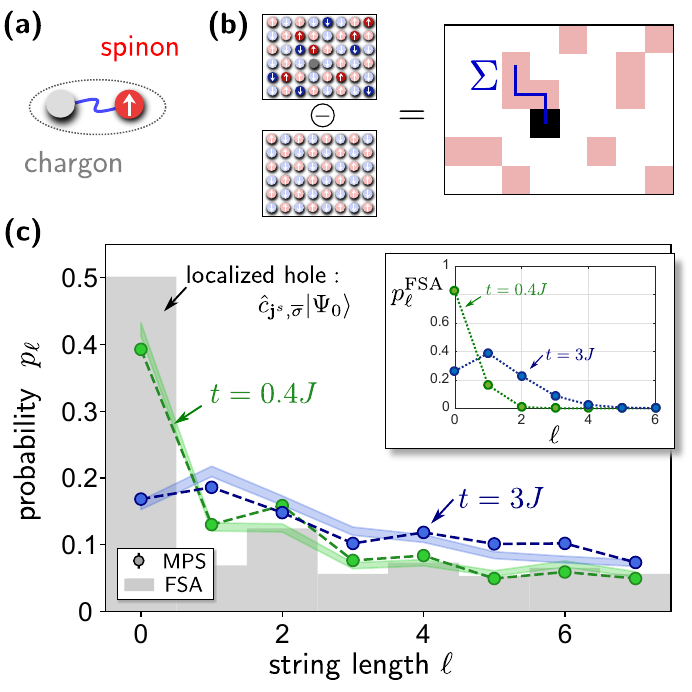, width=0.45\textwidth}~~~~
\caption{\textbf{Meson-like spinon-chargon bound states and short-range hidden string order.} A single hole in a 2D AFM forms a meson-like bound state of a spinon and a chargon, similar to quark-antiquark pairs forming mesons in high energy physics. (a) In the $t-J$ model, a spinon can be bound to a chargon by a geometric string of displaced spins. (b) Signatures of such strings ($\Sigma$) can be visualized in individual Fock configurations by analyzing the difference to a classical N\'eel pattern. We use the matrix product state formalism (MPS) and the DMRG algorithm to generate snapshots of the $T=0$ ground state of the $t-J$ model with a single hole, similar to the recent measurements using ultracold fermions \cite{Chiu2018}. In (c) we show the distribution function of the length $\ell$ of string-like patterns emanating from the hole. A striking difference is observed between a localized and a mobile hole (MPS, indicated by symbols connected with dashed lines). Mobile holes are described quantitatively by the geometric string theory (FSA, shaded ribbons) which is based on the string length distributions $p_\ell^{\rm FSA}$ shown in the inset.}
\label{figStrings}
\end{figure}

In this article, we demonstrate that the hidden string order paradigm discussed in Ref.~\cite{Chiu2018} provides a unified understanding of the properties of a single hole doped into an AFM parent state, with or without long-range magnetic order. We use numerical density-matrix renormalization group (DMRG) simulations on $8 \times 6$ cylinders to show that the hidden string order, for which recent experiments \cite{Chiu2018} have found indications at elevated temperatures, is also present in the ground state of a single hole in the $t-J$ model, see Fig.~\ref{figStrings}. This observation leads us to a microscopic description of individual holes doped into general spin backgrounds with strong AFM correlations. 

Specifically, we propose a trial wavefunction for the magnetic polaron formed by a single hole, which goes beyond the RVB paradigm by explicitly including short-range hidden string order. While most theories start from the weak coupling regime, where the tunneling rate $t$ of the hole is small compared to the super-exchange energy $J$, our method works best at strong couplings, where the bandwidth of a free fermion $W_t = 8 t$ is much larger than the energy range covered by the (para-) magnon spectrum $W_J \approx 2 J$, i.e. $t \gg J /4$. This coincides with the most relevant regime in high-temperature cuprate superconductors, where $t \approx 3 J$ \cite{Lee2006}. Note that we require $J/t \geq 0.05$ however, below which the Nagaoka polaron with a ferromagnetic dressing cloud is realized \cite{Nagaoka1966,White2001}. Extensions of our approach to weak couplings, $t \lesssim J$, are possible and will be devoted to future work. A central part of our study is the analysis of string patterns in individual images. This contains more information than the commonly used two point correlation functions and is motivated by recent experiments with quantum gas microscopes.

This paper is organized as follows. In the remainder of the introduction, we provide a brief review of the known properties of magnetic polarons along with an overview of our new results, concerning in particular the short-range hidden string order and the magnetic polaron radius. In Sec.~\ref{secModel} we discuss our microscopic model for describing individual dopants in an AFM and introduce the trial wavefunction. In Sec.~\ref{secResults} we present our numerical results and analyze the accuracy of the trial wavefunction. We close with an outlook and a discussion in Sec.~\ref{secDiscussion}.

\subsection{Magnetic polarons}
\label{subsebMagPolProp}
\begin{figure}[t!]
\centering
\epsfig{file=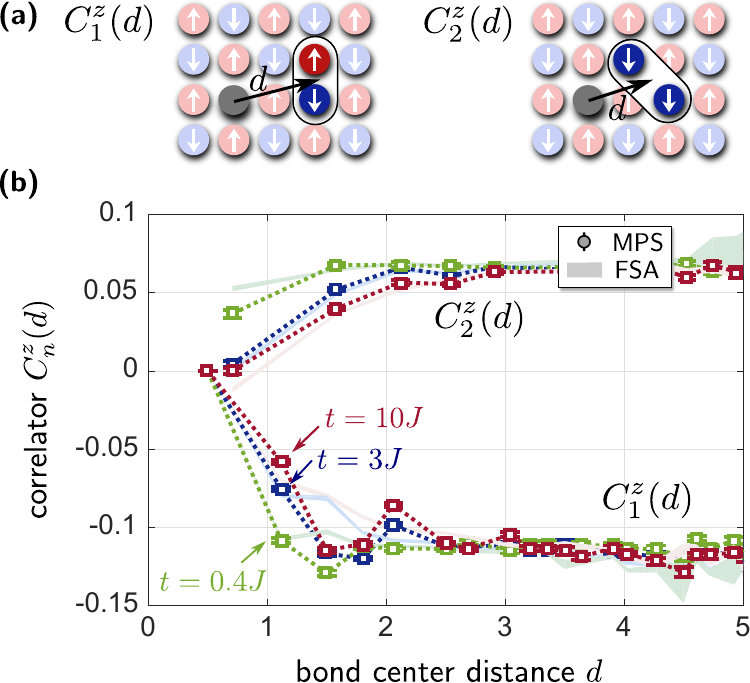, width=0.45\textwidth}~~~~
\caption{\textbf{Dressing cloud of magnetic polarons.} Using the MPS formalism and DMRG we calculate local spin correlation functions $C^z_n(d) = \langle \hat{n}^h_{\vec{r}_h} \hat{S}^z_{\vec{r}_2} \hat{S}^z_{\vec{r}_1} \rangle /  \langle \hat{n}^h_{\vec{r}_h} \rangle$, where $n=1$ ($n=2$) denotes nearest neighbor (diagonal next-nearest neighbor) configurations of $\vec{r}_1$, $\vec{r}_2$, as a function of the bond-center distance $d=|(\vec{r}_1 + \vec{r}_2 ) /2 - \vec{r}_h|$ of the two spins from the hole. At strong coupling the distortion of the magnetic spin environment around the hole follows an almost universal shape which is described well by the geometric string approach (FSA), except for some additional features between $d \approx 1.5$ and $d \approx 2.5$ captured only by DMRG. Calculations are performed for one hole and $S^z_{\rm tot}=1/2$ as described in Sec.~\ref{subsecStringParadigm}.}
\label{figMagPolEnv}
\end{figure}

When a single dopant is introduced into a spin background, it can be considered as a mobile impurity which becomes dressed by magnetic fluctuations and forms a new quasiparticle -- a magnetic polaron. In the case of a doublon or a hole doped into a Heisenberg AFM, commonly described by the $t-J$ model, the dressing by magnon fluctuations leads to strongly renormalized quasiparticle properties \cite{SchmittRink1988,Kane1989,Sachdev1989}. Here we provide a brief review of these known properties and their most common interpretation: 
\begin{itemize}
\item[(i)] The dispersion relation of the hole is strongly renormalized, with a bandwidth $W \propto J$ rather than the bare hole hopping $t$;
\item[(ii)] The shape of the dispersion differs drastically from that of a free hole, $-2 t [ \cos k_x + \cos k_y]$. It has a minimum at $\vec{k} = (\pi/2,\pi/2)$ and disperses weakly on the edge of the magnetic Brillouin zone (MBZ), $|k_x| + |k_y| = \pi$; see Fig.~\ref{figMPdispersion};
\item[(iii)] At strong couplings the ground state energy depends linearly on $J^{2/3} t^{1/3}$ and approaches $- 2 \sqrt{3} t$ when $J \to 0$.
\end{itemize} 
In the conventional magnetic polaron picture, (i) and (ii) are a consequence of a cloud of correlated magnons dressing the hole \cite{SchmittRink1988,Kane1989,Sachdev1989}, see Fig.~\ref{figMagPolEnv}. This polaron cloud is difficult to describe quantitatively due to the strong interactions of the magnons with the hole, with strength $t$. The properties in (iii) can be obtained from numerical calculations within the magnetic polaron theory, but their relation to an underlying physical mechanism is not made explicit. 

\subsection{Parton picture: Spinons, chargons and strings}
As reviewed next, the established properties of magnetic polarons (i) - (iii) follow more naturally from a spinon-chargon ansatz. Here the magnetic polaron is understood as a composite bound state of two partons:  A heavy spinon carrying the spin quantum number of the magnetic polaron, and a light chargon carrying its charge. This parton picture of magnetic polarons was first suggested by B\'eran et al. \cite{Beran1996}. Based on an even broader analysis these authors conjectured that magnetic polarons are composites, closely resembling pairs of quarks forming mesons in high-energy physics. 

The properties (i) and (ii) can be understood from the parton ansatz by noting that the chargon fluctuates strongly on a time scale $\sim 1/t$, whereas the center-of-mass motion of the spinon-chargon bound state is determined by the slower time scale $\sim 1/J$ of the spinon \cite{Beran1996}. Hence the overall kinetics of the bound state is dominated by the spinon dispersion, which features (i) a band-width $W \propto J$; and (ii) a near-degeneracy at the magnetic zone boundary because the spinon dynamics is driven by spin-exchange interactions on the bi-partite square lattice. 

As also previously recognized \cite{Bulaevskii1968,Brinkman1970,Trugman1988,Shraiman1988a,Manousakis2007,Golez2014}, the third result (iii) is related to the string picture of holes moving in a classical N\'eel state \cite{Chernyshev1999}: The hole motion creates strings of overturned spins, leading to a confining force \cite{Bulaevskii1968}. The approximately linear string tension $\propto J$ leads to the power-law dependence $\propto J^{2/3} t^{1/3}$ of the ground state energy, and the asymptotic value $-2 \sqrt{3} t$ is a consequence of the fractal structure of the Hilbert space defined by string states \cite{Bulaevskii1968,Brinkman1970,Grusdt2018PRX}. 

It is generally acknowledged that strings play a role for the overall energy of magnetic polarons in the $t-J$ model (iii), but the string picture alone does not account for the strongly renormalized dispersion relation of the hole, i.e. properties (i) and (ii) above. Hence a complete description of magnetic polarons needs to combine the parton ansatz with the string picture. As also noted by B\'eran et al. \cite{Beran1996}, it is natural to assume that the strings are responsible for binding spinons to chargons. 

While the combination of spinons, chargons and strings provides a satisfactory picture of magnetic polarons, a quantitative microscopic description of the meson-like bound states and the underlying partons has not been provided. More recently, toy models have been discussed which also contain spinon-chargon bound states from the start \cite{Punk2015PNASS} and capture the most important physical features of the pseudogap phase, see also Ref.~\cite{Baskaran2007}. But in these cases, too, the precise connection to the microscopic $t-J$ model and the spinon-chargon binding mechanism remain subject of debate.

In this article we introduce a complete microscopic description of magnetic polarons, in terms of individual spinons, chargons and so-called geometric strings \cite{Grusdt2018PRX,Grusdt2018mixD,Chiu2018} of displaced -- rather than overturned -- spins connecting the partons. This leads us to a new trial wavefunction of magnetic polarons, which can be constructed for arbitrary doped quantum AFMs, with or without long-range order. Our microscopic description implies new experimental signatures, which will be discussed next. They go beyond the capabilities of traditional solid state experiments, but can be accessed with ultracold atoms. 

This article extends our earlier work on the spinon-chargon theory for the simplified $t-J_z$ model with Ising couplings between the spins, where the spinon motion is introduced by Trugman loops \cite{Trugman1988,Grusdt2018PRX}, and in systems with mixed dimensionality where the hole motion is constrained to one direction \cite{Grusdt2018mixD}. Instead of invoking gauge fields for modeling the strings connecting spinons and chargons \cite{Beran1996}, our approach has a purely geometric origin and generalizes the concept of squeezed space used to describe doped 1D systems \cite{Ogata1990,Zaanen2001,Kruis2004a,Hilker2017}.

\subsection{Geometric paradigm: Short-range hidden string order in magnetic polarons}
\label{subsecStringParadigm}
The properties (i) and (ii) of magnetic polarons discussed in Sec.~\ref{subsebMagPolProp} have been measured in solid state experiments \cite{Zhou2008} by spectroscopic techniques. Quantum gas microscopy allows to go beyond such measurements and analyze individual experimental snapshots, obtained in quantum projective measurements, and directly search for signatures of string formation \cite{Bulaevskii1968,Brinkman1970,Trugman1988,Shraiman1988a,Manousakis2007,Golez2014} in real space \cite{Grusdt2018PRX}. This has been done in Ref.~\cite{Chiu2018}, where string patterns were analyzed and signatures for hidden AFM correlations have been obtained.

To identify string patterns, the experimental images in Ref.~\cite{Chiu2018} were compared to a perfect checkerboard configuration as expected for a classical N\'eel state. String like objects were revealed from difference images where the sites deviating from the perfect N\'eel pattern are identified, see Fig.~\ref{figStrings} (b). Then the distribution of lengths of such string-like objects was analyzed in Ref.~\cite{Chiu2018}. The string patterns exist even at half filling due to fluctuations of the local staggered magnetic moment, providing a background signal. Upon doping, a significant increase of the number of string patterns was detected which is proportional to the number of holes \cite{Chiu2018}.

Here we perform DMRG simulations on a $8 \times 6$ cylinder, using the TeNPy package \cite{Hauschild2018}, to search for similar signatures of geometric strings in the ground state of the $t-J$ model with exactly one hole, $\ket{\Psi_{\rm 1h}}$. To generate snapshots $\{ \ket{\alpha_n} \}$ of the wavefunction, we employ Metropolis Monte Carlo sampling of Fock basis states $\ket{\alpha}$ with one hole and calculate the required overlap $|\bra{\alpha} \Psi_{\rm 1h} \rangle|^2$ using the matrix product state formalism. For every generated snapshot $\ket{\alpha_n}$ we calculate the difference to a classical N\'eel pattern and determine the length $\ell$ of non-branching string-like defects emanating from the hole, see Fig.~\ref{figStrings} (b). Cases where no strings exist count as $\ell = 0$, and in cases with multiple strings the longest object is considered. Our analysis is similar to the experimental one in Ref.~\cite{Chiu2018}, except for the simultaneous spin and charge resolution which leads to a reduction of the background signal from undoped regions.

In Fig.~\ref{figStrings} (c) we show our results for the full counting statistics of string lengths $\ell$. The DMRG calculations were performed on a $8 \times 6$ cylinder with open (periodic) boundary conditions in the long (short) direction, in the sector with $\hat{S}^z_{\rm tot}=1/2$. We compare the case of a mobile hole at weak, $t=0.4 J$, and strong coupling, $t=3 J$, to a Heisenberg AFM with one spin removed, corresponding to a localized hole. In the latter case, the majority of strings has length $\ell = 0$. Other string lengths are also found, but even string lengths $\ell=2,4,...$ are more likely than odd ones. This is understood by noting that quantum fluctuations on top of the classical N\'eel state are caused by spin-exchange terms: individually, they change the string length by two units.  

\begin{figure}[t!]
\centering
\epsfig{file=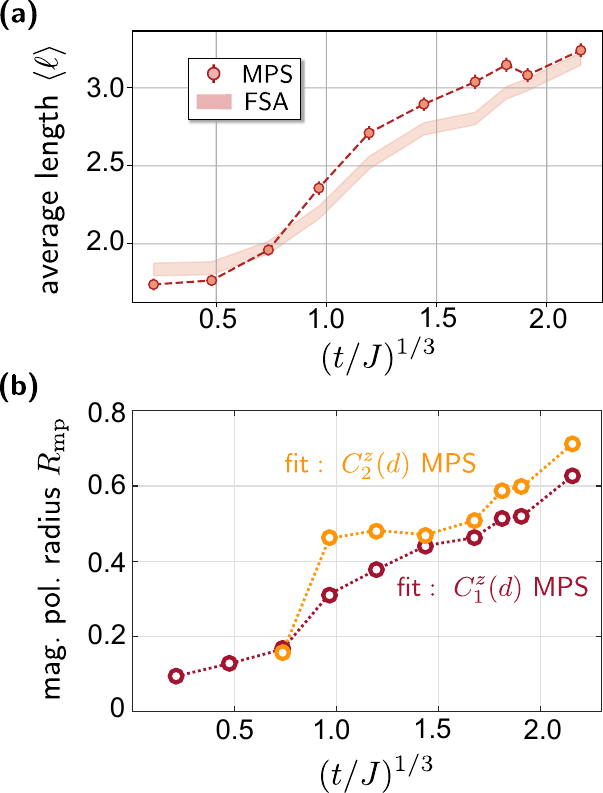, width=0.36\textwidth}~~~~~~~
\caption{\textbf{Magnetic polaron radius.} We calculate the size of the magnetic polaron as a function of $t/J$: (a) by determining the average length $\langle \ell \rangle$ of the string-like objects revealed in individual snapshots, and (b) by fitting the dependence of local spin correlations $C_n(d)$ on the bond center distance by a function of the form $C_n^\infty + a e^{- d / R_{\rm mp}}$ and interpreting the fit parameter $R_{\rm mp}$ as the magnetic polaron radius. For small values of $t$, the fit to $C_2^z(d)$ is not meaningful and we do not provide any data points in this regime.}
\label{figMagPolRad}
\end{figure}

We observe that the string length distributions obtained for mobile holes are significantly broader, and the even-odd effect explained by quantum fluctuations of the spins is much less pronounced. Already for $t \approx J$ we find that approximately half of the observed string patterns have a length $\ell \geq 2$. For $t=3 J$ the string length distribution continues to broaden and develops a local maximum at $\ell = 1$. In Fig.~\ref{figMagPolRad} (a) we plot the average string length, i.e. the first moment of the distribution. As expected from the linear string tension \cite{Bulaevskii1968}, it depends linearly on $(t/J)^{1/3}$ in the strong coupling regime.

The string patterns revealed in Fig.~\ref{figStrings} (c) in the ground state share the same characteristics as the string patterns found experimentally in Ref.~\cite{Chiu2018} over a wide range of dopings. As in the experiment, we will show in this article that the string patterns are described quantitatively by the so-called frozen spin approximation (FSA), which will be discussed in detail in Sec.~\ref{subsecFSA}. The main assumption of the FSA is to consider only charge fluctuations along strings of displaced spins. While the quantum state of the surrounding spins remains unmodified, the positions of the spins in the lattice change. 

This geometric paradigm is at the heart of the FSA and allows to construct a new set of snapshots $\{ \ket{\alpha^{\rm FSA}_n} \}$ describing a mobile hole, starting from a set of snapshots $\{ \ket{\alpha^{0}_n} \}$ for the undoped system: From the latter we generate the FSA snapshots by removing a spin in the center of the cylinder and moving the resulting hole in random directions $l$ times, where $l$ is sampled from the string length distribution $p_l^{\rm FSA}$ obtained from microscopic considerations, see Sec.~\ref{subsecFSA}. This motion of the hole displaces the spins and introduces short-range hidden string order. 

In Fig.~\ref{figStrings} (c) we analyze the string patterns in the FSA snapshots and find remarkable agreement with our full numerical DMRG simulations of the mobile holes. This remains true for a wide range of parameters $t/J$. Similar agreement was reported in ED simulations of a simplified $t-J$ model with mixed dimensionality \cite{Grusdt2018mixD}. The inset of Fig.~\ref{figStrings} (c) shows the underlying FSA string-length distributions $p_l^{\rm FSA}$, which share the same qualitative features as the detected string patterns in the main panel.

\subsection{Dressing cloud of magnetic polarons}
The capability of ultracold atom experiments to resolve the collapsed quantum state with full resolution of spin and charge simultaneously \cite{Boll2016} has recently lead to the first microscopic observation of the dressing cloud of a magnetic polaron \cite{Koepsell2018}. The measurements are consistent with earlier theoretical calculations at zero temperature \cite{Elser1990} and show that the local spin correlations are only affected in a relatively small radius of one to two lattice sites around the mobile dopant. In Fig.~\ref{figMagPolEnv} we perform similar calculations using DMRG, see also Ref.~\cite{White2001}, and observe that the spin correlations approach a universal shape which becomes nearly independent of $t/J$ at strong couplings, $t \gg J/4$. 

The most pronounced effect of the dopant is on the diagonal next-nearest neighbor correlations. They are strongly suppressed and can change sign for sufficiently large $t/J$ at finite temperature \cite{Salomon2018,Koepsell2018}. Such behavior can be understood from the FSA by noting that the charge is located at one end of the fluctuating geometric string, which interchanges the sub-lattice indices of the surrounding spins and hides the underlying AFM correlations. For $C_1$, $C_2$ denoting nearest and next-nearest neighbor spin correlations in the undoped system, the FSA predicts diagonal next-nearest neighbor correlators $C_2(d=1/\sqrt{2})$ directly next to the dopant given by
\begin{equation}
C_2 (1 / \sqrt{2}) |_{\rm FSA} \approx \l p_0^{\rm FSA} + \frac{1- p_0^{\rm FSA}}{2} \r C_2 + \frac{1- p_0^{\rm FSA}}{2} C_1.
\end{equation}
Here $d=1/\sqrt{2}$ denotes the bond-center distance defined in the caption of Fig.~\ref{figMagPolEnv}; $p_l^{\rm FSA}$ is the string-length distribution derived from the FSA approach in Sec.~\ref{subsecFSA}, which is shown in the inset of Fig.~\ref{figStrings} (c).

The correlations between the mobile hole and the surrounding spins are liquid like, with no significant effect on lattice sites more than two sites away even when $t/J$ is large: If we fit an exponential $C_n^\infty + a e^{- d / R_{\rm mp}}$ to $C_n(d)$, we find that  $R_{\rm mp}$ -- which we identify as the polaron radius -- depends only weakly on $t/J$ and the index $n$ of the correlator, see Fig.~\ref{figMagPolRad} (b). The average string length $\langle \ell \rangle$ of the string patterns revealed in individual microscopic Fock configurations, in contrast, depends more strongly on $t/J$, see Fig.~\ref{figMagPolRad} (a). 

In the string picture, these liquid like correlations in the local spin environment of the mobile dopant are a direct consequence of the large number of string configurations $N_\Sigma(\ell)$ with a specific string length $\ell$, growing exponentially: $N_\Sigma(\ell) =4 \times 3^{\ell-1}$ for $\ell > 0$. Every individual string configuration $\Sigma$ has a large effect on a specific set of spin correlations. But by averaging over all allowed string states, the effect on a specific spin correlator relative to the dopant is strongly reduced.

\section{Model}
\label{secModel}
We consider a class of 2D $t-J$ models with Hamiltonians of the form $\H_{t-J}=  \H_t + \H_J$, where
\begin{equation}
 \H_t = - t \sum_{\ij}  \sum_\sigma \hat{\mathcal{P}}_{\rm GW}  \bigl( \cd_{\vec{i},\sigma} \c_{\vec{j},\sigma} + \hc \bigr) \hat{\mathcal{P}}_{\rm GW}
\label{eqHt}
\end{equation}
describes tunneling of holes with amplitude $t$. We consider fermions $\c_{\vec{i},\sigma}$ with spin $\sigma$ and use Gutzwiller projectors $\hat{\mathcal{P}}_{\rm GW}$ to restrict ourselves to states with zero or one fermion per lattice site; $\ij$ denotes a pair of nearest neighbor (NN) sites and every bond is counted once in the sum. The second term, $\H_J$, includes interactions between the spins $\hat{\vec{S}}_{\vec{j}} = \sum_{\sigma,\tau = \uparrow, \downarrow} \cd_{\vec{j},\sigma} \vec{\sigma}_{\sigma,\tau} \c_{\vec{j},\tau}$ with an overall energy scale $J$. In the following we will consider NN Heisenberg exchange couplings, 
\begin{equation}
 \H_J =   J \sum_{\ij}  \l \hat{\vec{S}}_{\vec{i}} \cdot \hat{\vec{S}}_{\vec{j}} - \frac{\hat{n}_{\vec{i}} \hat{n}_{\vec{j}} }{4} \r
 \label{eqHJ}
\end{equation}
where $\hat{n}_{\vec{j}} = \sum_{\sigma = \uparrow,\downarrow} \cd_{\vec{j},\sigma} \c_{\vec{j},\sigma}$ denotes the number density of the fermions, but the methods introduced below can be applied more generally. 

To make the single-occupancy condition built into the $t-J$ model explicit, we use a parton representation,
\begin{equation}
\c_{\vec{j},\sigma} = \hd_{\vec{j}} \f_{\vec{j},\sigma}.
\label{eqDefSpnonHolon}
\end{equation}
Here $\h_{\vec{j}}$ is a bosonic chargon operator and $\f_{\vec{j},\sigma}$ is a $S=1/2$ fermionic spinon operator \cite{Auerbach1998,Wen2004}. The physical Hilbert space is defined by all states satisfying 
\begin{equation}
\sum_{\sigma} \fd_{\vec{j},\sigma} \f_{\vec{j},\sigma} + \hd_{\vec{j}} \h_{\vec{j}} = 1, \quad \forall \vec{j}.
\label{eqCondSingOcc}
\end{equation}

We start from the half-filled ground state $\ket{\Psi_0}$ of the undoped spin Hamiltonian $\H_J$, and consider cases where $\ket{\Psi_0}$ has strong AFM correlations. The ground state of the Heisenberg model Eq.~\eqref{eqHJ} has long-range AFM order, but the presence of strong and short-ranged AFM correlations would be sufficient to justify the approximations made below. We note that our results below do not require explicit knowledge of the wavefunction $\ket{\Psi_0}$.

The simplest state doped with a single hole is obtained by applying $\c_{\vec{j}^s,\overline{\sigma}} $ to $\ket{\Psi_0}$, where $\overline{\cdot}$ reverses the spin: $\overline{\uparrow} = \downarrow$, $\overline{\downarrow} = \uparrow$. This state,
\begin{equation}
\ket{\vec{j}^s,\sigma,0} = \c_{\vec{j}^s,\overline{\sigma}} \ket{\Psi_0} = \hd_{\vec{j}^s} \f_{\vec{j}^s,\overline{\sigma}} \ket{\Psi_0},
\label{eqDefStringVac}
\end{equation}
with a spinon and a chargon occupying the same lattice site $\vec{j}^s$, defines the starting point for our analysis of the parton bound state constituting the magnetic polaron. In the following we assume that $t \gg J$, which justifies a Born-Oppenheimer ansatz: first the initially created valence spinon at site $\vec{j}^s$ will be fixed and we determine the fast chargon fluctuations. Similar to nuclear physics, these fluctuations can involve virtual spinon anti-spinon pairs. In a second step we introduce the trial wavefunction as a superposition state of different valence spinon positions $\vec{j}^s$. Finally, we will derive the renormalized dispersion relation of the spinon-chargon bound state.

\subsection{Chargon fluctuations: Geometric strings and frozen spin approximation}
\label{subsecFSA}
\begin{figure}[t!]
\centering
\epsfig{file=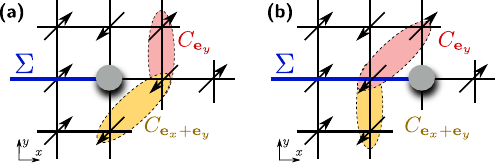, width=0.44\textwidth}
\caption{\textbf{Frozen spin approximation.} In the approximate FSA basis we only allow processes where the motion of the chargon displaces the surrounding spins without changing their quantum states. As a result, nearest neighbor correlations $C_{\vec{e}_y}$ -- red -- (next-nearest neighbor correlations $C_{\vec{e}_x+\vec{e}_y}$ -- yellow -- respectively) in the frozen spin background (a) contribute to next-nearest neighbor correlators (nearest neighbor correlators respectively) measured in states with longer string lengths (b). This leads to the approximately linear string tension, Eq.~\eqref{eqEffStringPot}, binding spinons to chargons.}
\label{figFSA}
\end{figure}

We review a binding mechanism of chargons and spinons by geometric strings, see also Sec. VI A in Ref.~\cite{Grusdt2018PRX}. When $t \gg J$, we expect that the chargon delocalizes until the energy cost for distorting the spin configuration around $\vec{j}^s$ matches the kinetic energy gain. To describe such chargon fluctuations we apply the "frozen-spin approximation" (FSA) \cite{Grusdt2018mixD,Chiu2018}: We assume that the motion of the hole merely displaces the surrounding spins, without changing their quantum state or their entanglement with the remaining spins. When the chargon moves along a trajectory $\mathcal{C}$ starting from the spinon, it leaves behind a string $\Sigma$ of displaced spins, which thus change their corresponding lattice sites. This so-called geometric string is defined by removing self-retracing paths from $\mathcal{C}$. For example, the spin operator located at site $\tilde{\vec{j}}$ initially, becomes $\tilde{\vec{S}}_{\tilde{\vec{j}}} = \hat{\vec{S}}_{\tilde{\vec{j}}-\vec{e}_x}$ when the chargon moves from $\tilde{\vec{j}}-\vec{e}_x$ to $\tilde{\vec{j}}$ along the string $\Sigma$, see Fig.~\ref{figFSA}.

The geometric string construction provides the desired generalization of squeezed space from 1D \cite{Ogata1990,Zaanen2001,Kruis2004a,Hilker2017} to 2D systems: the spins are labeled by their original lattice sites $\tilde{\vec{j}}$ before the chargon is allowed to move. We call this space, which excludes the lattice site $\vec{j}^s$ where the spinon is located, the 2D squeezed space. The motion of the chargon along a string $\Sigma$ changes the lattice geometry: the labels $\tilde{\vec{j}}$ no longer correspond to the actual lattice sites occupied by the spins. In particular, this changes the connectivity of the lattice, and spins which are NN in squeezed space can become next-nearest neighbors in real space. Hence, in this "geometric string" formulation the $\H_t$ part of the $t-J$ Hamiltonian is understood as introducing quantum fluctuations of the underlying lattice geometry. For an illustration, see Fig.~\ref{figFSA}.

When $t \gg J$, but before the Nagaoka regime is reached around $J/t \approx 0.05$, the spins in squeezed space do not have sufficient time to adjust to the fluctuating lattice geometry introduced by the chargon motion. We note that the shape and orientation of the geometric string are strongly fluctuating, which leads to spatial averaging of the effects of the string on the spins in squeezed space. This averaging is very efficient because the string is in a superposition of various possible configurations, the total number of which grows exponentially with the average string length. Hence the average effect on a given spin in squeezed space is strongly reduced, which provides a justification for the FSA ansatz. More technically, this means that the coupling of the fluctuating string to (para-) magnon excitations in squeezed space is weak and can be treated perturbatively.

Now we formalize our approach. When the chargon moves along a string $\Sigma$, starting from the state $\ket{\vec{j}^s,\sigma,0}$ in Eq.~\eqref{eqDefStringVac}, the many-body state within FSA becomes 
\begin{equation}
\ket{\vec{j}^s,\sigma,\Sigma} = \hat{G}_\Sigma \hd_{\vec{j}^s} \f_{\vec{j}^s,\overline{\sigma}} \ket{\Psi_0}.
\label{eqFSAmesonStates}
\end{equation}
Here the string operator, defined by
\begin{equation}
 \hat{G}_\Sigma = \prod_{\ij \in \Sigma} \bigg( \hd_{\vec{i}} \h_{\vec{j}} \sum_{\tau=\uparrow,\downarrow} \fd_{\vec{j},\tau} \f_{\vec{i},\tau} \bigg),
\end{equation}
creates the geometric string by displacing the spin states along $\Sigma$. The product $\prod_{\ij \in \Sigma}$ is taken over all links $\ij$ which are part of the string $\Sigma$, starting from the valence spinon position $\vec{j}^s$.

In a 2D classical N\'eel state, $\ket{\Psi_0^{\rm N}} = \ket{... \uparrow \downarrow \uparrow...}$, most string states $\ket{\vec{j}^s,\sigma,\Sigma}$ are mutually orthonormal. Specific configurations, so-called Trugman loops \cite{Trugman1988}, constitute an exception, but within an effective tight-binding theory it has been shown that this only causes a weak renormalization of the spinon dispersion \cite{Grusdt2018PRX}. Since the ground state $\ket{\Psi_0}$ of the infinite 2D Heisenberg model has strong AFM correlations, similar to a classical N\'eel state, we expect that the assumption that string states form an orthonormal basis remains justified. To check this, we calculated all such states with string lengths up to $\ell \leq 4$ and arbitrary spinon positions $\vec{j}^s$ using exact diagonalization (ED) in a $4 \times 4$ system. We found that $|\langle \vec{j}^s \! \! ~', \sigma, \Sigma' \ket{\vec{j}^s, \sigma, \Sigma}|^2 < 0.06$ unless $\Sigma = \Sigma'$ or $\Sigma$ and $\Sigma'$ are related by a Trugman loop. 

Now we will follow the example of Rokhsar and Kivelson: They introduced their celebrated dimer model \cite{Rokhsar1988} by defining a new basis which reflects the structure of the low-energy many-body Hilbert space in a class of microscopic spin systems. Similarly, we will postulate in the context of the FSA that all string states are mutually orthonormal. This defines the new basis of string states $\ket{\vec{j}^s, \sigma, \Sigma}$ which is at the heart of the FSA. Note, however, that we will return to the full physical Hilbert space of the original $t-J$ model later.

For a spinon with spin $\sigma$ fixed at $\vec{j}^s$, the effective string Hilbert space $\ket{\vec{j}^s,\sigma,\Sigma}$ has the structure of a Bethe lattice, or a Cayley graph. Its depth reflects the maximum length of the geometric string $\Sigma$, and the branches correspond to different directions of the individual string elements. The effective Hamiltonian $\H_{\rm eff}^t$ describing the chargon motion, i.e. fluctuations of the geometric string, consists of hopping matrix elements $t$ between neighboring sites of the Bethe lattice. In addition, the $J$-part $\H_J$ of the $t-J$ model gives rise to a potential energy term. Within the FSA, it can be easily evaluated,
\begin{multline}
\H_{\rm eff}^J = \sum_{\vec{j}^s,\sigma} \sum_\Sigma  \ket{\vec{j}^s,\sigma,\Sigma}  \bra{\vec{j}^s,\sigma,\Sigma} \\
\times  \underbrace{\bra{\vec{j}^s,\sigma,\Sigma} \H_J \ket{\vec{j}^s,\sigma,\Sigma}}_{V_{\rm pot}(\Sigma)}.
\label{eqHeffJFSA}
\end{multline}
Off-diagonal terms $\bra{\vec{j}^s \! ~',\sigma,\Sigma'} \H_J \ket{\vec{j}^s,\sigma,\Sigma}$ with $\vec{j}^s \! ~' \neq \vec{j}^s$ give rise to spinon dynamics and will be discussed below. By construction of the FSA, the potential $V_{\rm pot}(\Sigma)$ only depends on the spin-spin correlation functions in the undoped ground state, $C_{\vec{d}} = \bra{\Psi_0} \hat{\vec{S}}_{\vec{d}} \cdot \hat{\vec{S}}_{\vec{0}} \ket{\Psi_0}$.

We proceed as in the microscopic spinon-chargon theory of the $t-J_z$ model \cite{Grusdt2018PRX} and simplify the effective string Hamiltonian further by making the linear string approximation: We assume that the potential depends only on the string length $\ell_\Sigma$. From considering the case of straight strings, see Fig.~\ref{figFSA}, we obtain,
\begin{equation}
V_{\rm pot}(\Sigma) \approx \frac{dE}{d\ell} \ell_\Sigma + g_0 \delta_{\ell_\Sigma,0} + \mu_{\rm h},
\label{eqEffStringPot}
\end{equation}
with a linear string tension $dE/d\ell = 2 J (C_{\vec{e}_x+\vec{e}_y} - C_{\vec{e}_x} )$. The last term $\mu_{\rm h} = J (1 + C_{2 \vec{e}_x} - 5 C_{\vec{e}_x})$ corresponds to an overall energy offset; the middle term contributes only when the string length is $\ell_\Sigma = 0$ and describes a weak spinon-chargon attraction $g_0 = - J (C_{2\vec{e}_x} - C_{\vec{e}_x} )$. Note that we assumed four-fold rotational symmetry of $\ket{\Psi_0}$, e.g. $C_{\vec{e}_x} = C_{\vec{e}_y}$.

By solving the hopping problem on the Bethe lattice in the presence of the string potential \eqref{eqEffStringPot} as in Refs.~\cite{Bulaevskii1968,Brinkman1970,Shraiman1988a,Grusdt2018PRX}, we obtain approximations to the spinon-chargon binding energy $E_{\rm sc}^{\rm FSA}$ and the bound state wavefunction,
\begin{equation}
\ket{\psi_{\rm sc}^{\rm FSA}(\vec{j}^s,\sigma)} = \sum_\Sigma \psi_\Sigma^{\rm FSA} ~ \ket{\vec{j}^s,\sigma,\Sigma}.
\label{eqFSAshWvfct}
\end{equation}
Recall that we applied the strong coupling approximation, valid for $t \gg J$, and fixed the valence spinon at $\vec{j}^s$. Spinon dynamics will be discussed below. From Eq.~\eqref{eqFSAshWvfct} the FSA string length distribution is obtained, $p_l^{\rm FSA} = \sum_{\Sigma: \ell_\Sigma=l} |\psi_\Sigma^{\rm FSA}|^2$, which we used in Figs.~\ref{figStrings} - \ref{figMagPolRad}.

\subsection{Spinons at half filling}
Now we return to the analysis of spinons which determine the low-energy properties of magnetic polarons. We briefly review fermionic spinon representations of quantum AFMs and their corresponding variational wavefunctions. They provide the starting point for formulating a general spinon-chargon trial wavefunction in the next section.

For concreteness we consider the 2D Heisenberg Hamiltonian $\H_J$ in Eq.~\eqref{eqHJ} at half filling. Its ground state spontaneously breaks the ${\rm SU}(2)$ spin symmetry and has long-range N\'eel order \cite{Reger1988}. The corresponding low-energy excitations -- spin-$1$ magnons constituting the required Goldstone mode -- are most commonly described by a bosonic representation of spins, using e.g. Schwinger- or Holstein-Primakoff bosons. Recently it has been argued that the high-energy excitations of the AFM ground state can be captured more accurately by a fermionic spinon representation \cite{Piazza2015} however.
 
The fermionic spinon representation which we use in Eq.~\eqref{eqDefSpnonHolon} is partly motivated by analogy with the 1D $t-J$ model, where spinons can be understood as forming a weakly interacting Fermi sea \cite{Baskaran1987,Weng1995,Bohrdt2018}. On the other hand, Marston and Affleck \cite{Marston1989} have shown in 2D that the ground state of the Heisenberg model in the large-$N$ limit corresponds to the fermionic $\pi$-flux, or $d$-wave \cite{Baskaran1987}, state of spinons. For our case of interest, $N=2$, the $\pi$-flux state is not exact, but it can be used as a starting point for constructing more accurate variational wavefunctions. To this end we consider a general class of fermionic spinon mean-field states $\ket{\Psi_{\rm MF}(B_{\rm st},\Phi)}$, defined as the ground state at half filling of the following Hamiltonian, 
\begin{multline}
\H_{f, {\rm MF}} =  - J_{\rm eff} \sum_{\langle \vec{i}, \vec{j} \rangle, \sigma} \l  e^{i \theta^{\Phi}_{\vec{i},\vec{j}}} \fd_{\vec{j},\sigma} \f_{\vec{i},\sigma} + \hc \r \\
+ \frac{B_{\rm st}}{2} \sum_{\vec{j}, \sigma} (-1)^{j_x+j_y} \fd_{\vec{j},\sigma} (-1)^\sigma \f_{\vec{j},\sigma},
\label{eqDefHfMF}
\end{multline}
with Peierls phases $\theta^{\Phi}_{\vec{i},\vec{j}} = (-1)^{j_x+j_y+i_x+i_y} \Phi/4$ corresponding to a staggered magnetic flux $\pm \Phi$ per plaquette and a staggered Zeeman splitting $\sim B_{\rm st}$ which can be used to explicitly break the $SU(2)$ symmetry. 

A trial wavefunction for the $SU(2)$ symmetric $\pi$-flux state is obtained by applying the Gutzwiller projection \cite{Anderson1987,Baskaran1987} to the mean-field state with $\Phi= \pi$ and $B_{\rm st}=0$, i.e. $\ket{\Psi_\pi} = \hat{\mathcal{P}}_{\rm GW} \ket{\Psi_{\rm MF}(0, \pi)}$. Although it features no long-range AFM order, this trial state leads to a very low variational energy at half filling and it is also often considered as a candidate state at finite doping \cite{Lee2008a}. Another extreme is the $\Phi=0$ uniform RVB state with $B_{\rm st}=0$, i.e. $\ket{\Psi_0} = \hat{\mathcal{P}}_{\rm GW} \ket{\Psi_{\rm MF}(0, 0)}$, which also yields a very good variational energy at half filling. 

The best variational wavefunction of the general type $\hat{\mathcal{P}}_{\rm GW} \ket{\Psi_{\rm MF}(B_{\rm st}, \Phi)}$ has been found to have a non-zero staggered field $B_{\rm st} \neq 0$, consistent with the broken $SU(2)$ symmetry of the true ground state, and staggered flux $0 < \Phi < \pi$ \cite{Lee1988}. More recent calculations determined the optimal variational parameters of this "staggered-flux + N\'eel" (SF+N) trial state $\ket{\Psi_{\rm SF + N}} = \hat{\mathcal{P}}_{\rm GW}  \ket{\Psi_{\rm MF}(B^{\rm opt}_{\rm st}, \Phi^{\rm opt})}$ to be $\Phi^{\rm opt} \approx 0.4 \pi$ and $B^{\rm opt}_{\rm st} / J_{\rm eff} \approx 0.44$ \cite{Piazza2015}. The corresponding variational energy per particle $E_0^{\rm SF+N} / L^2 = -0.664 J$ is very close to the true ground state energy $E_0 / L^2 = -0.669 J$ known from first-principle Monte-Carlo simulations \cite{Trivedi1989}.

The main shortcoming of mean-field spinon theories as in Eq.~\eqref{eqDefHfMF} is that they neglect gauge fluctuations \cite{Wen2004}. These lead to spinon confinement in the ground state of the 2D Heisenberg model \cite{Wen2004} and, hence, free spinon excitations as described by Eq.~\eqref{eqDefHfMF} cannot exist individually. Indeed, if the Gutzwiller projection method is used to define a variational wavefunction, the underlying mean-field spinon dispersion is usually not considered to have a concrete physical meaning. We emphasize, however, that a single spinon can exist in combination with a chargon if they form a meson. In this case, which is of primary interest to us, we argue that the spinon dispersion \eqref{eqSFNspnonDisp} has a concrete physical meaning.

\begin{figure}[t!]
\centering
\epsfig{file=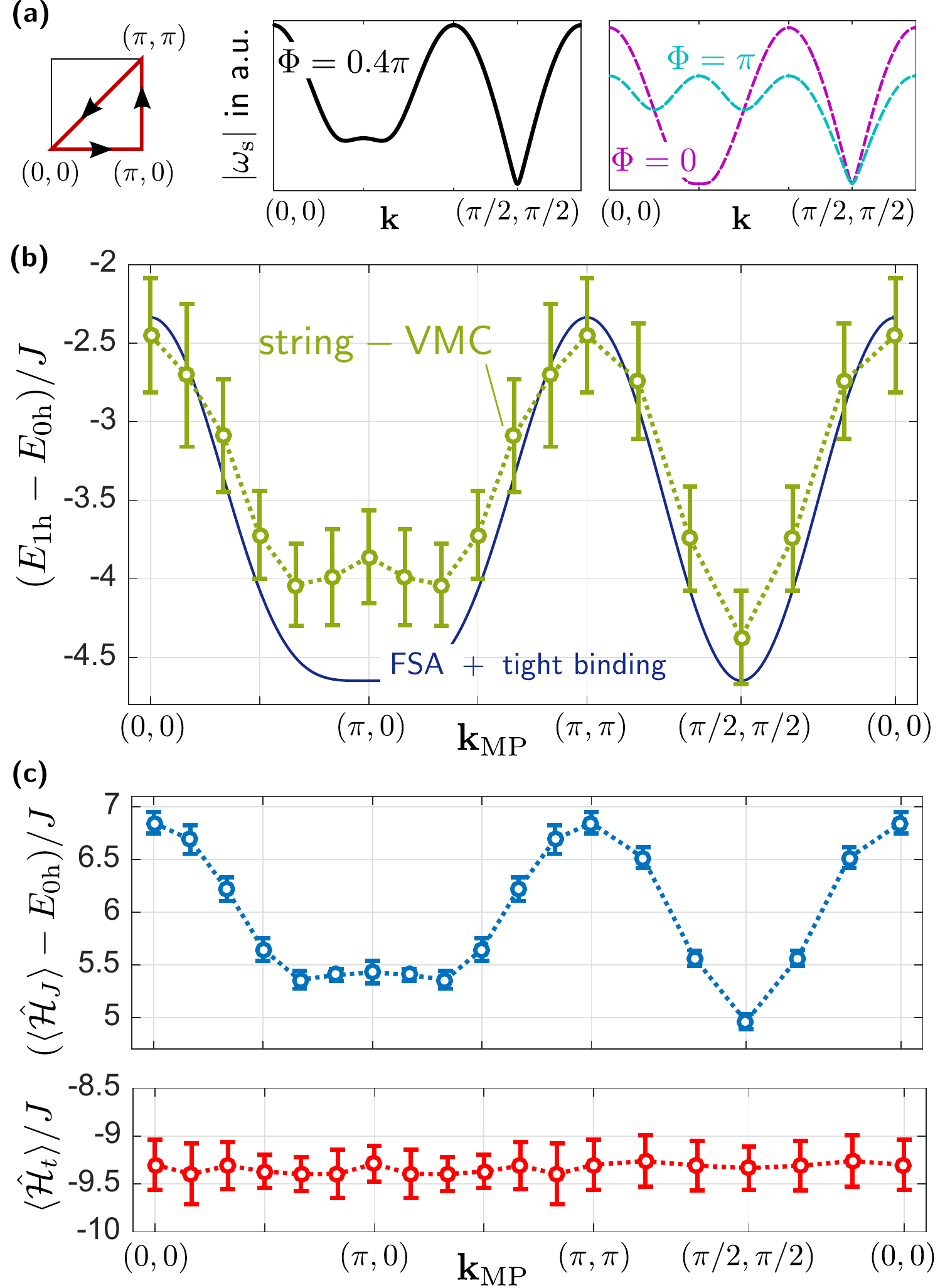, width=0.46\textwidth}~~
\caption{\textbf{Dispersion relation of a single hole in an AFM.} We consider cuts through the Brillouin zone along the path sketched in (a), starting at $(0,0)$. (a) The shape of the mean-field spinon dispersion, Eq.~\eqref{eqSFNspnonDisp}, is shown at $B_{\rm st}=0.44 J_{\rm eff}$ for $\Phi = 0.4 \pi$ (left) and $\Phi=0,\pi$ (right). (b) We calculate the magnetic polaron dispersion for $t=3J$ in a $12 \times 12$ system from the meson trial wavefunction, Eq.~\eqref{eqDefMesonWvfct}, (string-VMC) and compare it to predictions by the analytical FSA theory combined with a simplified tight-binding expression for the spinon dispersion, Eq.~\eqref{eqFSAmesonDisp}. (c) As expected from the strong coupling spinon-chargon picture, only the part of the variational energy associated with spin-exchange terms $\langle \H_J \rangle$ depends on $\vec{k}_{\rm MP}$ (top), whereas $\langle \H_t \rangle$ is non-dispersive (bottom).}
\label{figMPdispersion}
\end{figure}

The main difference between spinon models with different values of the staggered flux $\Phi$ is their dispersion relation. From the Hamiltonian Eq.~\eqref{eqDefHfMF} we obtain the mean-field spinon dispersion
\begin{equation}
\omega_{\rm s}(\vec{k}) = - \sqrt{4 J_{\rm eff}^2 \left| \cos(k_x) e^{-i \frac{\Phi}{4}} + \cos(k_y) e^{i \frac{\Phi}{4}} \right|^2 + \frac{B_{\rm st}^2}{4}}.
\label{eqSFNspnonDisp}
\end{equation}
For $B_{\rm st}=0$ and $\Phi \neq 0$ it has Dirac points at the nodal point $\vec{k}=(\pi/2,\pi/2)$. A finite staggered magnetic field $B_{\rm st}$ opens a gap everywhere. In this case the dispersion has a minimum at $\vec{k}=(\pi/2,\pi/2)$, unless $\Phi=0$ when the dispersion is degenerate along the edge of the magnetic zone boundary. The energy difference between the anti-nodal point $(0,\pi)$ and the nodal point $(\pi/2,\pi/2)$ is zero for $\Phi=0$ and maximal when $\Phi=\pi$, see Fig.~\ref{figMPdispersion} (a).

For the optimal variational parameters $\Phi^{\rm opt}$ and $B^{\rm opt}_{\rm st}$, the shape of the mean-field spinon dispersion relation \eqref{eqSFNspnonDisp} closely resembles the known dispersion of a single hole moving inside an AFM: It is weakly dispersive on the edge of the MBZ, has its minimum at $(\pi/2,\pi/2)$ and a pronounced maximum at $(0,0)$, see Fig.~\ref{figMPdispersion} (a). This is consistent with our conjecture from the spinon-chargon theory that the magnetic polaron dispersion is dominated by the spinon at strong couplings, $t \gg J$.

\subsection{Meson trial wavefunction}
To obtain a complete description of the meson-like bound state constituting a hole in an AFM, we combine geometric strings with the fermionic spinon representation. Starting from Eq.~\eqref{eqFSAmesonStates} with $\ket{\Psi_0} = \ket{\Psi_{\rm MF}^{\rm SF+N}} \equiv \ket{\Psi_{\rm MF}(B^{\rm opt}_{\rm st}, \Phi^{\rm opt})}$ we construct a translationally invariant trial wavefunction, 
\begin{equation}
\ket{\Psi_{\rm sc}(\vec{k}_{\rm MP})} = \sum_{\vec{j}^s}  \frac{e^{i \vec{k}_{\rm MP} \cdot \vec{j}^s}}{L}   \sum_\Sigma \psi_\Sigma ~ \hat{G}_\Sigma ~ \hat{\mathcal{P}}_{\rm GW} ~ \f_{\vec{j}^s,\overline{\sigma}} \ket{\Psi_{\rm MF}^{\rm SF+N}}.
\label{eqDefMesonWvfct}
\end{equation}
Here $\vec{k}_{\rm MP}$ denotes the total lattice momentum of the spinon-chargon magnetic polaron state, and $L$ denotes the linear system size. Note that we dropped the chargon operators $\h_{\vec{j}}$ in the expression because they are uniquely defined by the condition Eq.~\eqref{eqCondSingOcc} after the Gutzwiller projection in our case with single hole. 

In general, the values of the string wavefunction $\psi_\Sigma \in \mathbb{C}$ can be treated as variational parameters in Eq.~\eqref{eqDefMesonWvfct}, but in practice we use the result obtained explicitly from the FSA calculation in Eq.~\eqref{eqFSAshWvfct}: I.e. we set $\psi_\Sigma$ in Eq.~\eqref{eqDefMesonWvfct} equal to $\psi_\Sigma^{\rm FSA}$ determined in Eq.~\eqref{eqFSAshWvfct}. The undoped parent state $ \ket{\Psi_{\rm MF}^{\rm SF+N}}$ in Eq.~\eqref{eqDefMesonWvfct} can be replaced by any fermionic spinon mean-field wavefunction $\ket{\Psi^{f}_{\rm MF}}$. In particular, this allows to use Eq.~\eqref{eqDefMesonWvfct} to describe spinon-chargon bound states even in phases with deconfined spinon excitations.

We recapitulate the physics of Eq.~\eqref{eqDefMesonWvfct}: First, the valence spinon is created in the mean-field state. At strong couplings it carries the total momentum $\vec{k}_{\rm MP}$ of the meson-like bound state, $\f_{\vec{k}_{\rm MP},\overline{\sigma}} = L^{-1} \sum_{\vec{j}} e^{i \vec{k}_{\rm MP} \cdot \vec{j}} \f_{\vec{j},\overline{\sigma}}$. The Gutzwiller projection subsequently yields a state in the physical Hilbert space,
\begin{equation}
\hat{\mathcal{P}}_{\rm GW} \f_{\vec{k}_{\rm MP},\overline{\sigma}} \ket{\Psi_{\rm MF}^{\rm SF+N}} = \sum_{\vec{j}^s} \sum_{\alpha} \Phi_{\vec{k}_{\rm MP}}(\vec{j}^s,\alpha)  \hd_{\vec{j}^s} \f_{\vec{j}^s,\overline{\sigma}} \ket{\alpha}
\end{equation}
where $\sum_\alpha$ denotes a sum over all half-filled Fock states $\ket{\alpha}$. In this new state the spinon and chargon positions $\vec{j}^s$ coincide. In the last step we apply the string operators $\hat{G}_\Sigma$ to this state and create a superposition of fluctuating geometric strings in Eq.~\eqref{eqDefMesonWvfct}, which captures the internal structure of the meson-like bound state.

\subsection{Simplified tight-binding description of spinons}
\label{subsecTBspinons}
To obtain more qualitative analytical insights to the properties of meson-like spinon-chargon bound states, we return to the simplified FSA description developed in Sec.~\ref{subsecFSA} and extend it by an approximate tight-binding treatment of the spinon dispersion. This approach captures fewer details than the trial wavefunction Eq.~\eqref{eqDefMesonWvfct} but provides an intuitive physical picture of the main features revealed in the spinon dispersion.

\begin{figure}[t!]
\centering
\epsfig{file=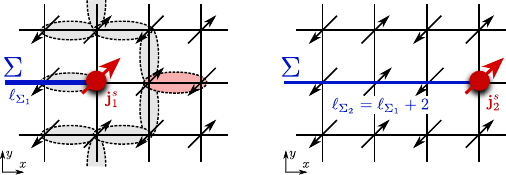, width=0.48\textwidth}
\caption{\textbf{Simplified tight-binding description of the spinon dispersion.} In a classical N\'eel state, a spinon at site $\vec{j}^s_1$ (left) can hop to another site $\vec{j}^s_2$ by spin-exchange processes on the bonds indicated by ellipses. The spin-exchange on the bond indicated by the red ellipse leads to the spinon configuration shown on the right, where $\vec{j}^s_2 = \vec{j}^s_1 + 2 \vec{e}_x$ and the string length changes by two units, $\ell_{\Sigma_2} = \ell_{\Sigma_1} + 2$. Similarly, spinons at $\vec{j}^s_2 = \vec{j}^s_1 - 2 \vec{e}_x$, $\pm 2 \vec{e}_y$, $+\vec{e}_x \pm \vec{e}_y$ and $-\vec{e}_x \pm \vec{e}_y$ can be reached, with  $\ell_{\Sigma_2} = \ell_{\Sigma_1} \pm 2$.}
\label{figTightBinding}
\end{figure}

So far the $J$-part of the effective FSA Hamiltonian, $\H_{\rm eff}^J$ in Eq.~\eqref{eqHeffJFSA}, includes only the string potential. Now we add terms $J_{\rm s}(\vec{j}^s_2,\vec{j}^s_1 ; \Sigma_2,\Sigma_1) \ket{ \vec{j}^s_2,\sigma,\Sigma_2 }\bra{\vec{j}^s_1,\sigma,\Sigma_1}$ changing the position of the valence spinon, with matrix elements 
\begin{equation}
J_{\rm s}(\vec{j}^s_2,\vec{j}^s_1 ; \Sigma_2,\Sigma_1) = \bra{\vec{j}^s_2,\sigma,\Sigma_2} \H_{J} \ket{\vec{j}^s_1,\sigma,\Sigma_1}.
\label{eqDefJs}
\end{equation}
We first evaluate Eq.~\eqref{eqDefJs} for string states $\ket{\vec{j}^s,\sigma,\Sigma}$ constructed from a classical N\'eel state $\ket{\Psi_0} = \ket{\Psi_0^{\rm N}}$. Ignoring loop configurations of the strings $\Sigma_{1,2}$ \cite{Trugman1988}, one obtains non-zero matrix elements $J_{\rm s} = J/2$ only if $\vec{j}^s_1$ and $\vec{j}^s_2$ can be connected by two links in arbitrary directions and if the string length changes by two units, $\ell_{\Sigma_2} = \ell_{\Sigma_1} \pm 2$; such pairs of sites are shown in Fig.~\ref{figTightBinding} and we will denote them as $\langle \langle \vec{j}^s_2, \vec{j}^s_1 \rangle \rangle$. 

Next we check that these terms remain dominant when the string states are constructed from the exact ground state $\ket{\Psi_0}$ of the 2D Heisenberg model. To this end we performed ED in a $4\times 4$ system with periodic boundary conditions and confirmed that the matrix elements between sites $\langle \langle \vec{j}^s_2, \vec{j}^s_1 \rangle \rangle$ remain dominant with magnitudes $J_{\rm s}  =0.52 J$ close to the result from the classical N\'eel state. In contrast, matrix elements \eqref{eqDefJs} between states with spinons on neighboring sites $\langle \vec{j}^s_2, \vec{j}^s_1 \rangle$ remain small, $|J_{\rm s}(\vec{j}^s_2,\vec{j}^s_1)| <0.1 J$, and will be neglected.

In a generic quantum AFM we expect that this picture remains valid, at least qualitatively. The spin-exchange couplings on the bonds around the spinon indicated in Fig.~\ref{figTightBinding} can be written as $\hat{\vec{S}}_{\vec{i}} \cdot \hat{\vec{S}}_{\vec{j}} = \hat{P}_{\vec{i},\vec{j}} / 2 - 1/4$, where $\hat{P}_{\vec{i},\vec{j}} \ket{\sigma_{\vec{i}}, \sigma_{\vec{j}}} = \ket{\sigma_{\vec{j}}, \sigma_{\vec{i}}}$ exchanges the two spins irrespective of their orientation. Hence both spins change their sublattice index, which is expected to lead to a large overlap with a state describing a geometric string of length $\ell_{\Sigma_2} = \ell_{\Sigma_1} \pm 2$ and a modified spinon position. The corresponding matrix element is thus expected to be $J_s \approx J/2$, irrespective of the details of the undoped AFM $\ket{\Psi_0}$.

Now we focus on strong couplings, $t \gg J$, and make a Born-Oppenheimer ansatz \cite{Bruderer2007}. As in Eq.~\eqref{eqFSAshWvfct} we first fix the spinon position and determine the ground state of the fluctuating geometric string, $\ket{\psi_{\rm sc}^{\rm FSA}(\vec{j}^s,\sigma)}$. Here we have to be careful in order to avoid double counting: In our original derivation of the FSA string potential in Eq.~\eqref{eqEffStringPot} we included the energy $J \langle \hat{\vec{S}}_{\vec{i}} \cdot \hat{\vec{S}}_{\vec{j}} \rangle$ between any pair of spins on sites $\ij$ as a constant energy. However, we argued above that the exchange part $J \hat{P}_{\vec{i}, \vec{j}} /2$ of the Heisenberg couplings $J \hat{\vec{S}}_{\vec{i}} \cdot \hat{\vec{S}}_{\vec{j}}$ on the bonds shown in Fig.~\ref{figTightBinding} lead to spinon dynamics, and we will include them in the tight-binding spinon Hamiltonian. Hence, to avoid double counting of terms in $\H_J$, we modify the effective string potential in Eq.~\eqref{eqEffStringPot} by subtracting $J \langle \hat{P}_{\vec{i}, \vec{j}} \rangle /2 =  J \bra{\Psi_0} \hat{\vec{S}}_{\vec{i}} \cdot \hat{\vec{S}}_{\vec{j}} + \nicefrac{1}{4} \ket{\Psi_0}$ for bonds $\ij$ contributing to the matrix elements in Eq.~\eqref{eqDefJs}. 

Next we include spinon dynamics. Due to the presence of geometric strings, the spinon hopping elements $J_{\rm s}$ between sites $\langle \langle \vec{j}^s_2, \vec{j}^s_1 \rangle \rangle$ are renormalized by a Franck-Condon overlap $\nu_{\rm FC}$,
\begin{equation}
J_{\rm s}^* = \sum_{\Sigma_1, \Sigma_2} J_{\rm s}(\vec{j}^s_2,\vec{j}^s_1;  \Sigma_2,\Sigma_1) ~ (\psi_{\Sigma_2}^{\rm FSA})^* \psi_{\Sigma_1}^{\rm FSA} = \nu_{\rm FC} ~ J_{\rm s},
\end{equation}
where $\psi_\Sigma^{\rm FSA}$ denotes the FSA string wavefunction from Eq.~\eqref{eqFSAshWvfct}. The resulting tight-binding hopping Hamiltonian gives rise to the strong coupling expression for the spinon-chargon energy,
\begin{multline}
E_{\rm sc}(\vec{k}) = 2 J_{\rm s} ~\nu_{\rm FC} \bigl[ 2 \cos(k_x+k_y) + 2 \cos(k_x-k_y) \\
+ \cos(2 k_x) + \cos(2 k_y)  \bigr] + E_{\rm sc}^{\rm FSA},
\label{eqFSAmesonDisp}
\end{multline}
where $E_{\rm sc}^{\rm FSA}$ is the energy contribution from the fluctuating geometric string.

The Franck-Condon factor can be calculated in the limits $t / J \to \infty, 0$. For weak couplings $t \ll J$ the string length becomes short and the Franck-Condon factor approaches zero, $\nu_{\rm FC} \to 0$. This leads to a strong suppression of the magnetic polaron bandwidth $W$. For strong couplings, $t \gg J$, the Franck-Condon factor approaches $\nu_{\rm FC} = 1/2$. This leads to a bandwidth $W \propto J \ll t$.

\section{Results}
\label{secResults}
Now we present numerical results from the microscopic spinon-chargon theory, obtained from the simplified tight-binding description, see Sec.~\ref{subsecTBspinons}, and from the variational energy of the trial wavefunction in Eq.~\eqref{eqDefMesonWvfct}. To calculate the latter we utilize standard Metropolis sampling, commonly employed in variational Monte Carlo (VMC) calculations, see e.g. Ref.~\cite{Gros1989}.

\subsection{Dispersion relation}
\emph{Shape.--}
In Fig.~\ref{figMPdispersion} (b) we show the variational dispersion relation, $E_{\rm MP}(\vec{k}_{\rm MP}) =E_{\rm 1h}(\vec{k}_{\rm MP}) - E_{\rm 0h}$ of the single hole in the AFM, or magnetic polaron, where $E_{\rm 0h} = E_0^{\rm SF+N}$ is the variational energy of the SF+N state without doping and $E_{\rm 1h}(\vec{k}_{\rm MP}) = \bra{ \Psi_{\rm sc}(\vec{k}_{\rm MP}) } \H_{t-J} \ket{ \Psi_{\rm sc}(\vec{k}_{\rm MP})}$ for doping with one hole. The result is in good quantitative agreement with numerical Monte Carlo calculations \cite{Liu1992,Brunner2000}, capturing all properties of the single-hole dispersion.

From the spinon-chargon theory we expect that the dispersion of a hole in an AFM is dominated by the spinon properties when $t \gg J$. Indeed, the shape of the variational dispersion in Fig.~\ref{figMPdispersion} (b) closely resembles the mean-field spinon dispersion, see Fig.~\ref{figMPdispersion} (a). To corroborate this picture further, we calculate the variational energies $\langle \H_J \rangle$ and $\langle \H_t \rangle$ as a function of $\vec{k}_{\rm MP}$ individually in Fig.~\ref{figMPdispersion} (c). Only the spin-exchange part $\langle \H_J \rangle$ is dispersive, whereas the chargon part $\langle \H_t \rangle$ does not depend on $\vec{k}_{\rm MP}$ within error bars. This is a direct indication that a hole in an AFM has two constituents, one of which mainly affects spin exchanges and determines the dispersion of the meson-like bound state. 

In Fig.~\ref{figMPdispersion} (b) we also compare our result from the spinon-chargon trial wavefunction to the FSA tight-binding prediction from Eq.~\eqref{eqFSAmesonDisp}. While there is remarkable overall agreement, the semi-analytical FSA tight-binding calculation misses some important qualitative features: Eq.~\eqref{eqFSAmesonDisp} does not capture the minimum of the single-hole dispersion at $(\pi/2,\pi/2)$ but predicts a degenerate minimum along the edge of the MBZ, resembling more closely the mean-field spinon dispersion \eqref{eqSFNspnonDisp} without staggered flux, $\Phi=0$, see Fig.~\ref{figMPdispersion} (a).

\begin{figure}[t!]
\centering
\epsfig{file=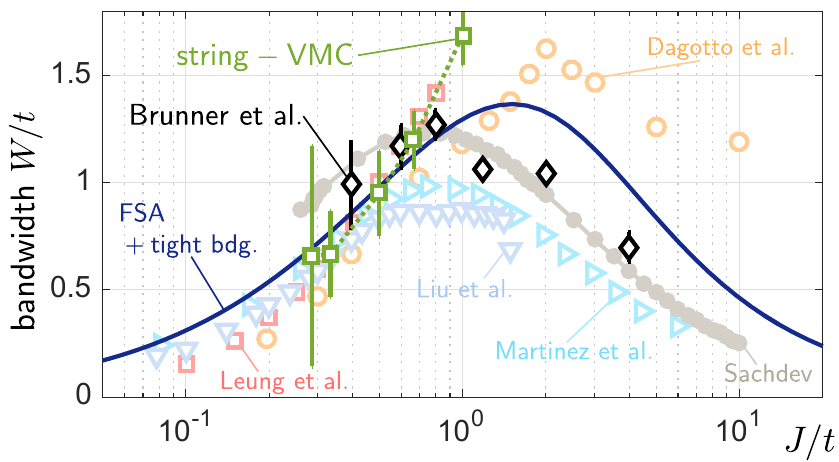, width=0.46\textwidth} ~~~
\caption{\textbf{Bandwidth $W$ of a single hole in an AFM.} We compare our variational results (string-VMC) from Eq.~\eqref{eqDefMesonWvfct}, valid at strong couplings, to quantum Monte Carlo simulations by Brunner et al.~\cite{Brunner2000}, ED studies by Dagotto et al.~\cite{Dagotto1990} and Leung and Gooding \cite{Leung1995}, a variational wavefunction by Sachdev \cite{Sachdev1989}, and spin-wave calculations by Martinez et al.\cite{Martinez1991} and by Liu and Manousakis \cite{Liu1992}. The overall shape of $W(J/t)$ is well captured by the effective FSA theory, Eq.~\eqref{eqFSAmesonDisp}. The string-VMC calculations are performed at $B_{\rm st}=0.44 J_{\rm eff}$, $\Phi = 0.4 \pi$ in a $12 \times 12$ system.}
\label{figMPbandWidth}
\end{figure}

\emph{Bandwidth.--}
In Fig.~\ref{figMPbandWidth} we vary the ratio $J/t$ and calculate the bandwidth $W = E_{\rm MP}(0,0) - E_{\rm MP}(\pi/2,\pi/2)$. At strong couplings, $t \gg J$, our results from the trial wavefunction \eqref{eqDefMesonWvfct} (string-VMC) are in good agreement with numerical results from various theoretical approaches. When $ t \gtrsim J/3$, the errorbars of our variational results are large. In this regime the average length of the geometric string exceeds one lattice site, and our string-VMC calculations suffer from strongly fluctuating numerical signs. 

We also compare the bandwidth $W$ of the magnetic polaron to the tight-binding prediction in Eq.~\eqref{eqFSAmesonDisp}. While quantitative agreement is not achieved everywhere, the overall dependence on $J/t$ is accurately captured. At strong couplings, $t \gg J$, the bandwidth $W \propto J$ is proportional to $J$, i.e. strongly suppressed relative to $t$. At weak couplings, $t \ll J$, the Franck-Condon factor vanishes, $\nu_{\rm FC} \to 0$, which leads to a strong suppression of $W$ relative to both $t$ and $J$.

\subsection{Ground state energy}
\begin{figure}[t!]
\centering
\epsfig{file=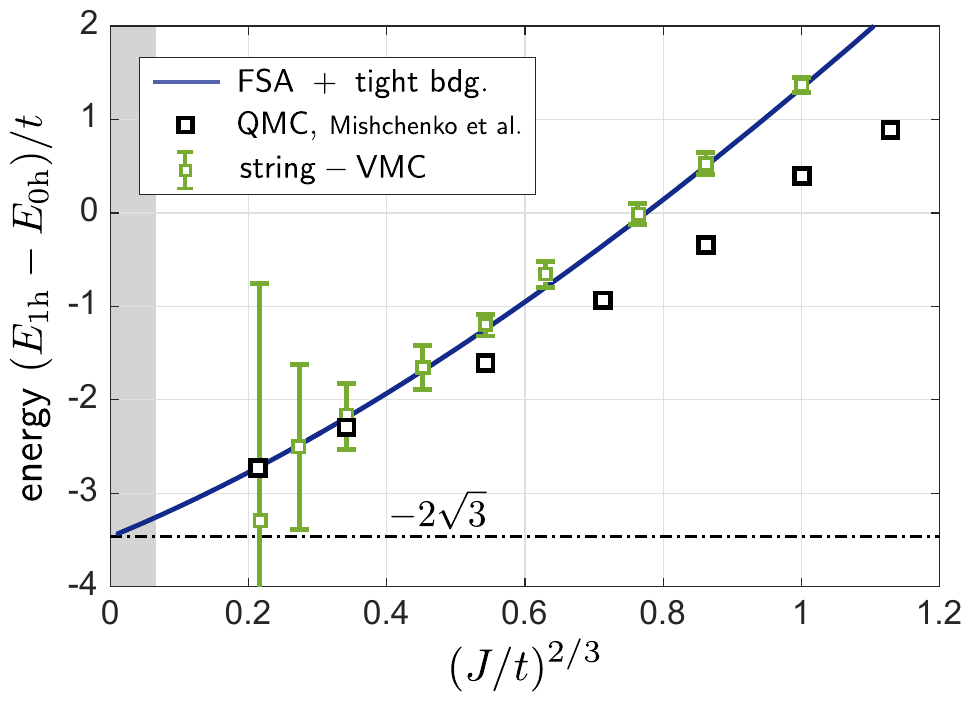, width=0.46\textwidth}~~
\caption{\textbf{Ground state energy of a hole in an AFM.} We compare our variational result from the meson trial wavefunction in Eq.~\eqref{eqDefMesonWvfct} (string-VMC) to quantum Monte Carlo calculations by Mishchenko et al.~\cite{Mishchenko2001} (QMC) and semi-analytical predictions by the tight-binding FSA theory from Eq.~\eqref{eqFSAmesonDisp}. In the shaded region, defined by $J < 0.05 t$, the ground state is expected to be a Nagaoka polaron \cite{Nagaoka1966,White2001}. The string-VMC calculations are performed at $B_{\rm st}=0.44 J_{\rm eff}$, $\Phi = 0.4 \pi$ in a $12 \times 12$ system at $\vec{k}_{\rm MP}=(\pi/2,\pi/2)$.}
\label{figMPenergy}
\end{figure}

In Fig.~\ref{figMPenergy} we calculate the ground state energy of the meson-like bound state as a function of $J/t$. The variational result (string-VMC) from the trial wavefunction \eqref{eqDefMesonWvfct} agrees well with numerically exact Monte Carlo calculations \cite{Mishchenko2001} at strong couplings, before the fluctuating signs prevents efficient numerical calculation. The deviations from the exact result are on the order of $J$ even when $J>t$. When $t \gg J$, the ground state energy of the single hole is of the form $E_{\rm MP} = - 2 \sqrt{3} t + c~  t^{1/3} J^{2/3} + \mathcal{O}(J)$, which can be understood as a consequence of the geometric string with an approximately linear string tension  \cite{Bulaevskii1968,Grusdt2018PRX}. The variational ground state energy is accurately described by the semi-analytical FSA prediction from Eq.~\eqref{eqFSAmesonDisp}, for all values of $t/J$. 

So far we evaluated the spinon-chargon trial wavefunction \eqref{eqDefMesonWvfct} using the FSA string wavefunction and set $\psi_\Sigma=\psi_\Sigma^{\rm FSA}$, where $\psi_\Sigma^{\rm FSA}$ was obtained from Eq.~\eqref{eqFSAshWvfct}. Because the number of allowed string states $\Sigma$ grows exponentially with the maximum length of the strings, it is numerically too costly to treat all amplitudes $\psi_\Sigma$ as variational parameters. To study the quality of the trial state, we now introduce a single variational parameter. We calculate the string wavefunction $\psi_\Sigma$ in the over-complete FSA Hilbert space but modify the potential in Eq.~\eqref{eqEffStringPot} by rescaling the linear string tension $dE/d\ell \to \lambda_{dE/d\ell} ~ dE/d\ell$. The numerical factor $\lambda_{dE/d\ell} \geq 0$ is then used as a variational parameter which controls the average string length.

\begin{figure}[t!]
\centering
\epsfig{file=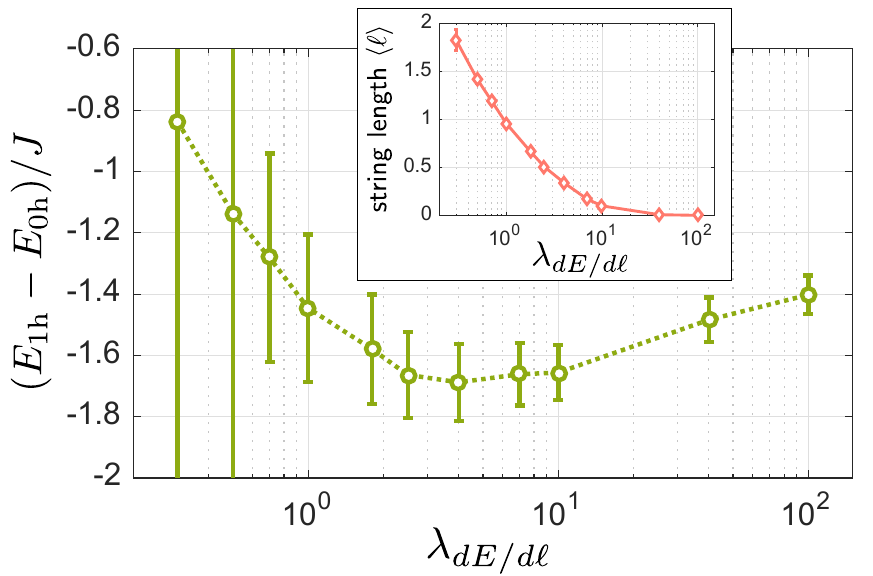, width=0.46\textwidth} ~~
\caption{\textbf{Optimization of the trial wavefunction.} We change the average length of the geometric string in the spinon-chargon wavefunction Eq.~\eqref{eqDefMesonWvfct}, $\langle \ell \rangle = \sum_\Sigma |\psi_\Sigma|^2 \ell_\Sigma$ shown in the inset, by rescaling the linear string tension in Eq.~\eqref{eqEffStringPot} with a factor $\lambda_{dE/d\ell}$. The resulting variational energy $\langle \H_{t-J} \rangle - E_{\rm 0h}$, in units of $J$, is calculated as a function of $\lambda_{dE/d\ell}$ for the parameter $t =2 J$. The string-VMC calculations are performed at $B_{\rm st}=0.44 J_{\rm eff}$, $\Phi = 0.4 \pi$ in a $12 \times 12$ system at $\vec{k}_{\rm MP}=(\pi/2,\pi/2)$.}
\label{figMPstringLength}
\end{figure}

In Fig.~\ref{figMPstringLength} we show the variational energy as a function of $\lambda_{dE/d\ell}$. We observe a minimum at approximately $\lambda_{dE/d\ell} \approx 4$, around which the variational energy depends rather insensitively on $\lambda_{dE/d\ell}$. For larger values of $\lambda_{dE/d\ell}$, where the average length of the geometric string is close to zero, higher variational energies are obtained. This indicates that the formation of geometric strings is energetically favorable. For smaller values of $\lambda_{dE/d\ell}$ the average string length exceeds one lattice constant. This makes the fluctuations of the numerical sign in the VMC method worse, but our results indicate an increase of the variational energy in this regime as well. In combination, these results support the spinon-chargon pairing mechanism by geometric strings.

\subsection{Magnetic polaron cloud}
\begin{figure}[b!]
\centering
\epsfig{file=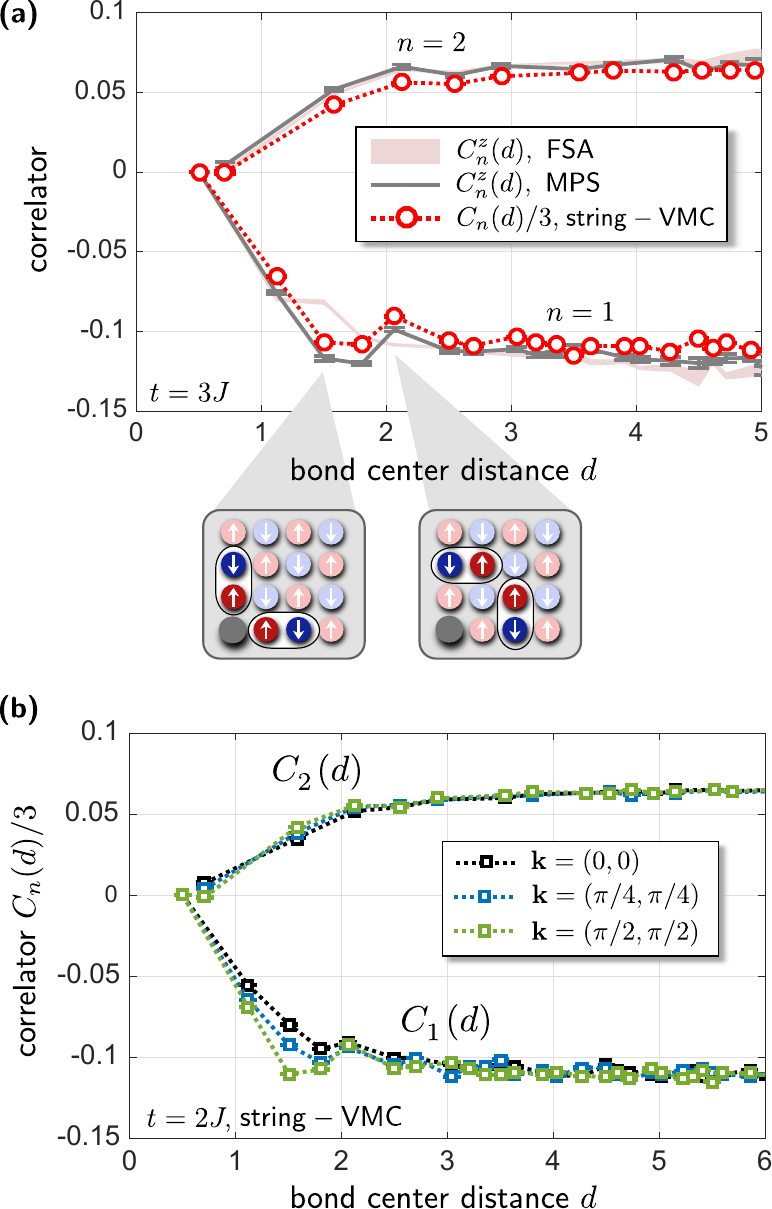, width=0.46\textwidth}~~
\caption{\textbf{Magnetic polaron cloud.} Using the trial wavefunction (string-VMC) in Eq.~\eqref{eqDefMesonWvfct} we calculate the local spin correlations $C_n(d) = \langle \hat{n}^h_{\vec{r}_h} \hat{\vec{S}}_{\vec{r}_1} \cdot \hat{\vec{S}}_{\vec{r}_2} \rangle /  \langle \hat{n}^h_{\vec{r}_h} \rangle$, where $d=|(\vec{r}_1 + \vec{r}_2 ) /2 - \vec{r}_h|$ is the bond-center distance between $\vec{r}_1$ and $\vec{r}_2$ and $n=1$ ($n=2$) corresponds to nearest (next-nearest) neighbor spin correlations. (a) For $t=3J$ we compare our string-VMC result $C_n(d)/3$ to $C^z_n(d)$ obtained from the same DMRG simulations as in Fig.~\ref{figMagPolEnv}: MPS corresponds to the full solution in the case of a mobile hole and the FSA predictions are generated from snapshots of the undoped Heisenberg AFM. (b) We calculate the momentum dependence of $C_n(d)$ from the trial wavefunction at $t=2J$. The string-VMC calculations in (a) and (b) are performed at $B_{\rm st}=0.44 J_{\rm eff}$, $\Phi = 0.4 \pi$ in a $14 \times 14$ system; in (a) $\vec{k}=(\pi/2,\pi/2)$.}
\label{figMPcloud}
\end{figure}

Finally we study the dressing cloud of magnetic polarons and calculate local spin correlations from the trial wavefunction \eqref{eqDefMesonWvfct}. In Fig.~\ref{figMPcloud} (a) we compare the variational result to our DMRG simulations described in Sec.~\ref{subsecStringParadigm} for $t=3J$. The DMRG results are based on the same snapshots from which we obtained the string length histogram in Fig.~\ref{figStrings} (c). This method also allows us to compare to predictions by the FSA, see Fig.~\ref{figMagPolEnv}, where geometric strings are included by hand into snapshots of an undoped Heisenberg model, see Sec.~\ref{subsecStringParadigm}.

For $t=3J$, i.e. for strong couplings, we find excellent agreement of the trial wavefunction with DMRG simulations. The numerical results confirm that diagonal next-nearest neighbor correlations next to the mobile hole are strongly suppressed, i.e. $C_2(1/\sqrt{2}) \approx 0$. At distances $d \gtrsim 2.5$, no significant dependence of the correlations on $t/J$ can be identified by either method. 

\begin{figure*}[t!]
\centering
\epsfig{file=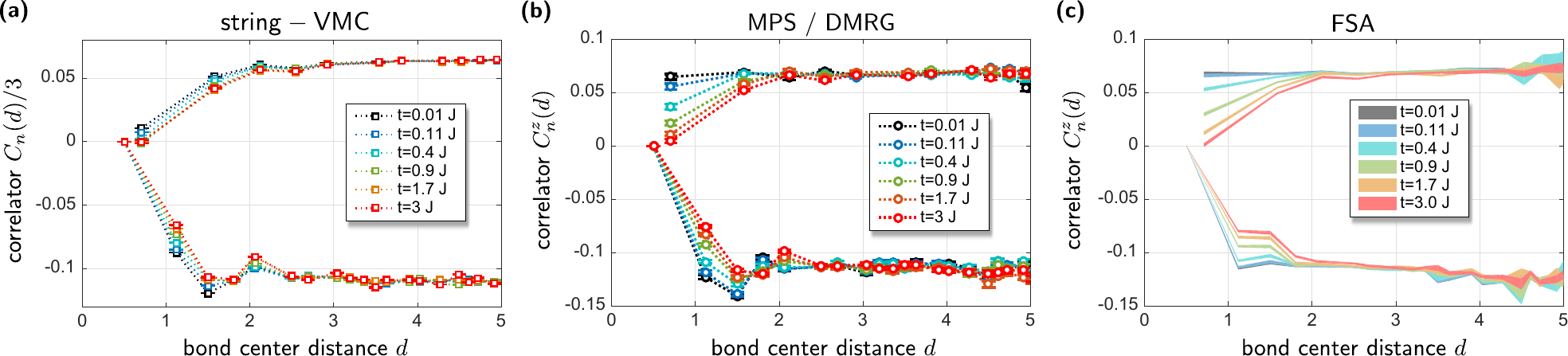, width=0.99\textwidth} 
\caption{\textbf{Comparison of the different theoretical approaches.} We calculate the local spin correlations $C_n(d)/3$ and $C_n^z(d)$ for $n=1,2$ in the vicinity of a mobile dopant, using different theoretical approaches: (a) the trial wavefunction (string-VMC) from Eq.~\eqref{eqDefMesonWvfct}, (b) by numerical DMRG simulations (MPS) and (c) using the FSA and starting from snapshots for the undoped Heisenberg model generated by DMRG. The string-VMC calculations in (a) are performed at $B_{\rm st}=0.44 J_{\rm eff}$, $\Phi = 0.4 \pi$ in a $14 \times 14$ system at $\vec{k}=(\pi/2,\pi/2)$.}
\label{figMPcloudComparison}
\end{figure*}

The most striking feature predicted by DMRG and the trial wavefunction is the formation of a peak with reduced nearest neighbor correlations $C_1(d)$ at the distance $d=2.06$ from the mobile dopant. This feature becomes more pronounced as $t/J$ increases. Additionally, we observe enhanced nearest neighbor correlations $C_1(d)$ at $d=1.5$, but this feature disappears for values of $t/J \gtrsim 0.5$, see Fig.~\ref{figMPcloudComparison} (a) and (b).

The additional spatial structure featured by the trial wavefunction and DMRG at $d=1.5$ and $d=2.06$ is not captured by the simplified FSA approach, see Fig.~\ref{figMPcloudComparison} (c). The fact that the feature is present in the trial wavefunction when $t \ll J$ indicates that it is caused by the microscopic correlations of the spinon position with its spin environment. The emergence of a second length scale, in addition to the string length $\ell \propto (t/J)^{1/3}$ captured by the FSA, can be considered as an indirect indication of fermionic spinon statistics: The Fermi momentum $k_{\rm F}$ defines a second intrinsic length scale in this case.

For weaker couplings, $t \lesssim J$, the trial wavefunction is less accurate since the Born-Oppenheimer approximation is no longer valid. As shown in Fig.~\ref{figMPcloudComparison} (a), it predicts a strong suppression of diagonal next-nearest neighbor correlations around the hole, $C_2(1/\sqrt{2}) \approx 0$, for all values of $t/J$. In contrast, the DMRG features a strong dependence of $C_2(1/\sqrt{2})$ on $t/J$ when $t \lesssim J$, which is accurately described by the FSA, see Fig.~\ref{figMPcloudComparison} (b) and (c). On the other hand, the FSA approach is less reliable for the nearest neighbor correlations, whose qualitative shape is remarkably well described by the trial wavefunction for all values of $t/J$.

In Fig.~\ref{figMPcloud} (b) we calculate local spin-spin correlations at a distance $d$ from the hole, for different total momenta $\vec{k}$ of the spinon-chargon bound state. We observe that the feature in $C_1(d)$ at $d=2.06$ is most pronounced in the ground state at $\vec{k}=(\pi/2,\pi/2)$. For $\vec{k}=(0,0)$ the additional structure disappears almost completely. This observation supports our earlier conclusion that the characteristic structure of $C_1(d)$ is related to the spinon, which also carries the center-of-mass momentum $\vec{k}$ of the magnetic polaron at strong couplings.

\section{Discussion and outlook}
\label{secDiscussion}
We have introduced a microscopic theoretical framework to describe a hole doped into an AFM as a meson-like bound state of a spinon and a chargon. As a binding mechanism of spinons and chargons at strong couplings, we suggest geometric strings: They model how the chargon motion dynamically changes the underlying lattice geometry. The trial wavefunction introduced in this article puts earlier results \cite{Beran1996} on a mathematical footing, allowing quantitative predictions beyond the simplified $t-J_z$ model \cite{Grusdt2018PRX}. While we focus on a single hole at zero temperature in a spin system with long-range AFM order, our method should also be applicable at finite doping, for higher temperatures and in systems without long-range N\'eel order. 

Our results obtained here for the energy, the dispersion relation and the magnetic dressing cloud of a single hole in an AFM are in good agreement with the commonly used magnetic polaron theory \cite{SchmittRink1988,Kane1989,Sachdev1989,Martinez1991,Liu1992}. The spinon-chargon approach can be understood as a refinement of the magnetic polaron picture. We find that many properties of the polarons formed by holes in the Fermi Hubbard model follow more directly from the spinon-chargon ansatz. Moreover, simple theoretical pictures can be derived from the parton approach at strong couplings, where the magnetic polaron theory becomes notoriously difficult to solve, and our approach is particularly useful for understanding the structure of magnetic polarons in real space, which has recently become accessible by quantum gas microscopy of ultracold atoms in optical lattices \cite{Koepsell2018}. 

Another key advantage of the spinon-chargon approach is that it continuously connects systems with and without long-range AFM order. In contrast to the magnetic polaron theory, our variational wavefunction captures correctly the physics of the 1D $t-J$ model, where geometric strings become infinitely long and spinons are no longer bound to chargons. We thus believe that our method is well suited to study the dimensional cross-over from the 1D to the 2D $t-J$ model in the future. In this article we only considered the case when spinons and chargons form a bound state, but we do not exclude the possibility that the attractive potential between spinons and chargons is finite and an unbound state could exist at finite energy. This would correspond to a phase with deconfined spinons.

As an important application, we expect that our approach can also provide the means for a microscopic description, starting form first principles, of the ${\rm FL}^*$ state proposed as an explanation of the pseudogap phase in cuprates \cite{Punk2015}. As a key ingredient, the ${\rm FL}^*$ state contains bound states of spinons and chargons, similar to the meson-like bound states discussed in this paper. We propose geometric strings as a possible spinon-chargon binding mechanism in this finite-doping regime. Indeed, recent experiments in the corresponding region of the cuprate phase diagram \cite{Chiu2018} have found indications for the presence of geometric strings, and here we confirmed these results at low doping and for zero temperature by state-of-the-art DMRG simulations. 

To shed more light on the connection between the spinon-chargon trial wavefunction and the pseudogap phase, as a next step it would be useful to study spectral properties of mesons as measured in angle-resolved photo emission spectroscopy (ARPES) experiments. From the strong coupling wavefunction introduced here, we expect two contributions to the spectral weight: a spinon part, which is strongly dispersive, and a chargon, or string, contribution which only has a weak momentum dependence. We expect that this allows to draw further analogies with ARPES spectra in 1D systems \cite{Weng1995,Bohrdt2018}, and it may provide new insights to the physics of Fermi arcs observed in the pseudogap phase of cuprates. 

We close by a comment about the relation of our approach to Anderson's resonating valence bond picture of high-temperature superconductivity \cite{Anderson1987,Anderson2004}. As in his approach, we use Gutzwiller projected mean-field states of spinons \cite{Baskaran1987} as key ingredients in our trial wavefunction. By adding geometric strings we include short-range hidden order and take Anderson's ansatz in a new direction. Our method is not based on spin-charge separation but instead describes meson-like bound states of spinons and chargons. The implications for unconventional superconductivity will be explored in the future.\\

\section*{Acknowledgements}
We would like to thank E. Altman, M. Knap, M. Punk, S. Sachdev, T. Shi, R. Verresen, Y. Wang and Z. Zhu for useful feedback and comments. We also acknowledge fruitful discussions with I. Bloch, C. Chiu, D. Chowdhury, D. Greif, M. Greiner, C. Gross, T. Hilker, S. Huber, G. Ji, J. Koepsell, S. Manousakis, F. Pollmann, A. Rosch, G. Salomon, U. Schollw\"ock, L. Vidmar, J. Vijayan and M. Xu. 

F.G. acknowledges support by the Gordon and Betty Moore foundation under the EPIQS program. F.G. and A.B. acknowledge support from the Technical University of Munich - Institute for Advanced Study, funded by the German Excellence Initiative and the European Union FP7 under grant agreement 291763, from the DFG grant No. KN 1254/1-1, and DFG TRR80 (Project F8). A.B. also acknowledges support from the Studienstiftung des deutschen Volkes. E.D. and F.G. acknowledge support from Harvard-MIT CUA, NSF Grant No. DMR-1308435,
AFOSR Quantum Simulation MURI.

\newpage

\section*{References}
\vspace{-1cm}

\end{document}